# Data Silos – A Roadblock for AIOps

**Subhadip, Kumar**\*

Western Governors University

**Abstract:** Using artificial intelligence to manage IT operations, also known as AIOps, is a trend that has attracted a lot of interest and anticipation in recent years. The challenge in IT operations is to run steady-state operations without disruption as well as support agility" can be rephrased as "IT operations face the challenge of maintaining steady-state operations while also supporting agility [11]. AIOps assists in bridging the gap between the demand for IT operations and the ability of humans to meet that demand. However, it is not easy to apply AIOps in current organizational settings. Data Centralization is a major obstacle for adopting AIOps, according to a recent survey by Cisco [1]. The survey, which involved 8,161 senior business leaders from organizations with more than 500 employees, found that 81% of them acknowledged that their data was scattered across different silos within their organizations. This paper illustrates the topic of data silos, their causes, consequences, and solutions.

## 1 INTRODUCTION

AIOps, or Artificial Intelligence for IT Operations, is a new approach that leverages machine learning and automation to enhance the observability and reliability of complex software systems. Observability is the ability to monitor and understand the internal state of a system or application based on the external outputs, such as logs, metrics, and traces. However, observability alone is not enough to cope with the increasing volume, velocity, and variety of data generated by modern applications. AIOps can help analyze and act on the data collected by observability tools and provide critical insights and recommendations for improving the performance, security, and user experience of the system. It also aggregates, normalizes and enriches events collected from fragmented tools, and uses AI to correlate that data into actionable insights.

AIOps is especially useful for large scale organizations that consist of several different applications, ranging from data lakes to enterprise solutions, monolith applications to microservices, on-premises to cloud native applications. A disruption of any of these applications can have a range of effects, such as flight delay, timely shipment of live saving medicines, revenue loss, and customer dissatisfaction. At the same time, it is almost impossible to manually monitor all these related applications and take corrective actions. That's where AIOps can help. By leveraging the data generated by these applications, such as logs, signals, traces and metrics, AIOps can filter out signals from noises, quickly learn and deliver critical insights, identify anomalies and outliers, perform performance analysis, augment decision making, conduct potential root cause analysis and finally in some cases enable automatic recovery from the failure.

However, to achieve these benefits, AIOps requires a centralized data source, also known as a source of truth, that collects, stores, and integrates data from multiple sources. Without right data, AIOps model will generate noises, create frustration among SREs and developers and the entire goal will fail. A centralized data source provides a consistent and accurate view of the system's state and behavior, as well as the relationships and dependencies among its components. It enables AIOps to apply advanced analytics and automation across the entire data pipeline, from ingestion to visualization. This way, AIOps can deliver actionable insights and recommendations that help SREs and developers achieve full stack observability and improve the quality and efficiency of their software delivery.

On the other side Data silos prevent AIOps from accessing and utilizing the data it needs to perform its functions. Data silos create data duplication, inconsistency, and incompleteness, which reduce the data quality and accuracy. Data silos also create data latency, unavailability, and irrelevance, which decrease the data timeliness and usefulness, it also generates incomprehensibility, inaccessibility, and insufficiency, which impair the data analysis and actionability. Without right data, AIOps model will generate noises, create frustration among SREs and developers and the entire goal will fail.

This article will address these challenges and show how an enterprise can overcome data silos to enable and accelerate the AIOps journey.

---

\* Place the footnote text for the author (if applicable) here.

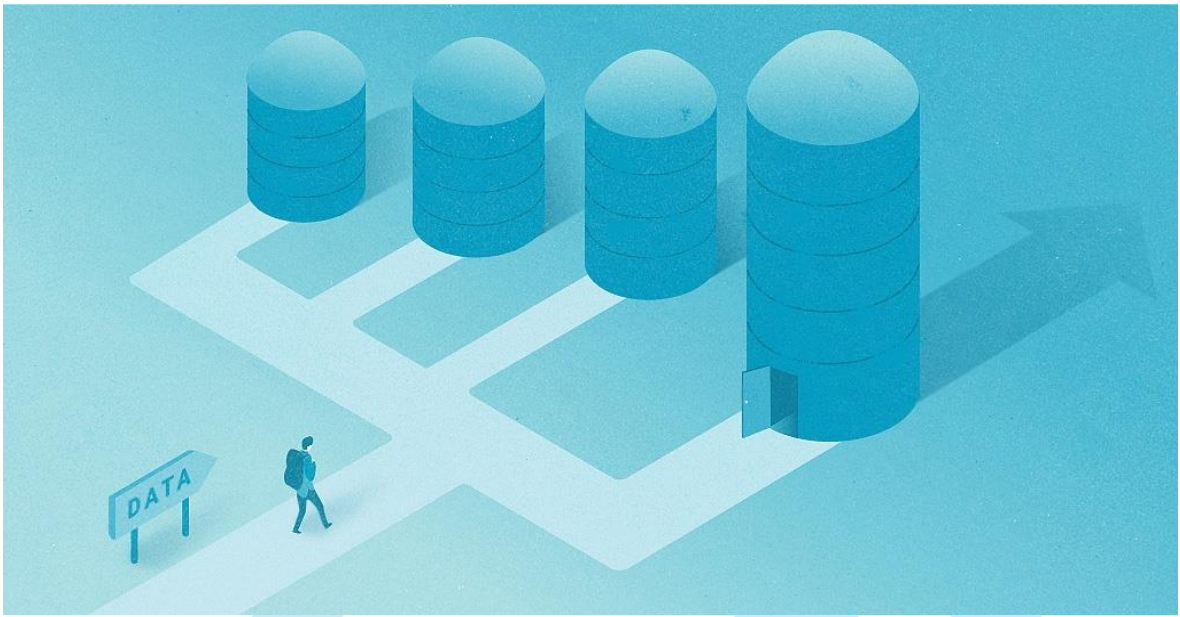

Fig 1. AIOps methodologies: Data, insights, and actions [15]

## 2   WHAT ARE DATA SILOS?

In the age of artificial intelligence, data science, machine learning, and predictive analytics, generating insights from the right data at the right time is critical [2]. Data or information silos are isolated data repositories that only one group in an organization can access or use effectively. Different departments inside IT organizations often keep their data in different locations which often referred as data silos. The name comes from the farm buildings that store different kinds of crops. Data silos tend to increase as more and more data is generated and stored in various formats. As a result, businesses collect a lot of data but use very little of it, because most of the data is isolated and unorganized.

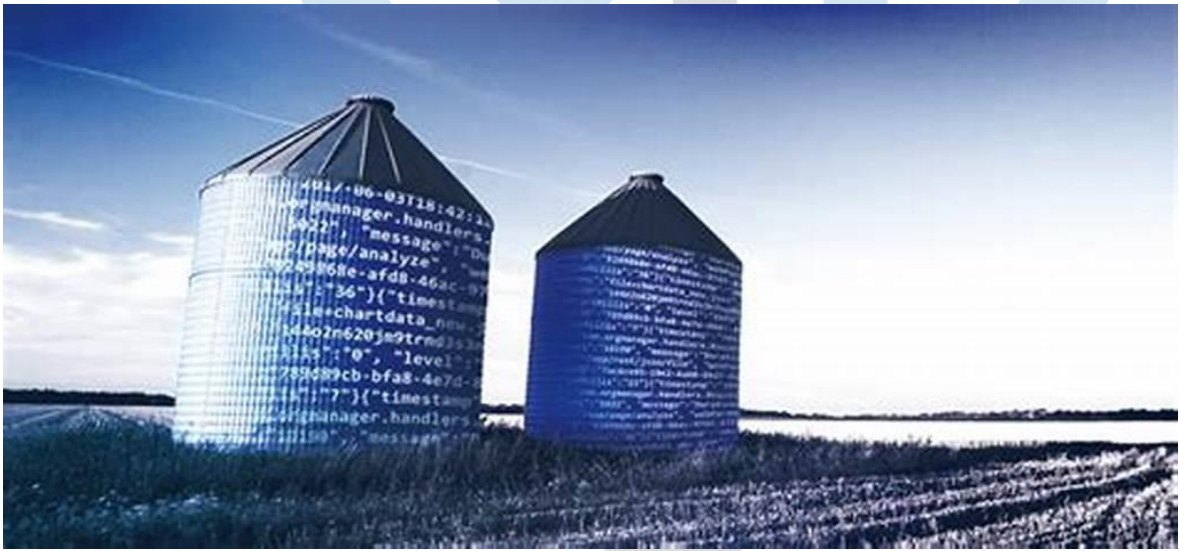

Fig 2. Farm building – data silos analogy

Data silos have existed for a long time in organizations, but their effects have become more significant in the modern era. Data silos are the result of storing and managing data in separate and isolated systems that are not easily accessible or shared by other groups or departments. This leads to inefficiencies, inconsistencies, and missed opportunities for data analysis and insights. Data silos also hinder collaboration and communication among teams and stakeholders, as well as innovation and growth. In today's world, where data is abundant and diverse, and where data-driven decision making is crucial, data silos pose a serious challenge for organizations that want to leverage their data assets and gain a competitive edge.

## 3 WHY THE DATA SILOS OCCUR?

One of the causes of data silos in organizations is the adoption of new technologies over time based on business requirements. Sometimes, data is copied from one system to another to perform calculations or analysis on it. However, this creates duplicate data sets that are stored in different systems that do not communicate with each other. This leads to data silos, which are isolated and inaccessible data repositories that prevent data sharing and integration. Priorities changes over years – 20 years back the priority was to generate a faster revenue report, now it is to perform to apply machine learning models on the revenue based on other related data.

Some says these silos aren't necessarily anyone's fault — it's just a natural consequence of time, growth, and evolution in a business [3]. Other says data silos happen due to a lack of forward planning [9]. Here are few possible reasons:

- Incompatible data formats and standards: Data silos can occur when different applications use incompatible formats or standards for their data. This prevents them from sharing or accessing data from each other. For example, one application may use XML while another may use JSON. This creates a barrier for data integration and exchange. Sometimes, another application may need to pull data from these applications for a specific purpose. However, if the data formats or standards are not compatible, this may not be possible. This leads to data silos that limit the availability and usability of data across the organization.
- Legacy system and technology: Legacy systems are outdated systems that are still in use by the organizations. Often such systems are hard to replace with modern software due to the amount of customization and integration that is in place and many a times due to legal requirements like data residency. They often pose a challenge for creating a centralized data source, which is a single and consistent repository of data that can be accessed and used by different applications and users. Legacy systems may have different data formats, structures, standards, and quality than modern systems. They may also have limited or no compatibility or interoperability with other systems or platforms. This makes it difficult to integrate, migrate, or synchronize data from legacy systems to a centralized data source. For example, a legacy system may use a flat file or a hierarchical database, while a modern system may use a relational or a NoSQL database. These systems have different data models, schemas, and queries that require complex and costly data transformation and mapping processes. Legacy systems may also have security, performance, and reliability issues that can affect the quality and availability of data. Therefore, legacy systems are often a barrier to a centralized data source that can enable data-driven decision making and innovation.
- Organization culture and politics: Organization silos often leads to data silos. They are interrelated and mutually reinforcing. Organization silos create data silos by limiting the access, availability, and quality of data for different groups or functions. Data silos, in turn, reinforce organization silos by creating information asymmetry, distrust, and inefficiency among different groups or functions. For example, in one large organization infrastructure and operation department do not communicate with application team. They both adopted their own APM tool without checking with each other. Both the tool generates the same OS metrics as CPU/Memory and Disk utilizations and stores/analyze by two different tool and in two different formats. This leads to inconsistent, inaccurate, or incomplete data that cannot easily be integrated or analyzed across the organization. This can also lead to conflicting, contradictory, or competitive behaviors that undermine the overall organizational objectives and outcomes. Cultural issues within the company can also play a big role in silos forming due to inter-office politics, interdepartmental politics, and team cliques. This can lead to an unfriendly work environment with information hoarding, decreasing productivity and efficiency [10].
- Lack of data retention policies: Data retention policy is a set of rules and guidelines that determine how long data should be stored and when it should be deleted or archived. This is part of data governance. Many large enterprises lack a defined data retention policy, which can lead to data accumulation and duplication from different sources over the years. This can result in data that is inaccurate, inconsistent, and irrelevant for AIOps. AIOps relies on high-quality and relevant data to perform tasks such as pattern recognition, anomaly detection, root cause analysis, and incident resolution. However, if the data is not properly managed and retained, AIOps may produce inaccurate and confused results that can affect the performance and reliability of IT systems and services. Therefore, having a defined data retention policy is essential for ensuring the quality and usefulness of data for AIOps.

## 4 IMPACT OF DATA SILOS

These are some of the impact of data silos.

- Wasted resources and poor productivity: AIOps tools heavily rely on single source of truth. Now when same data being deduplicated at different places, AIOps tools are unable to generate a meaningful insight. Data silos can also hinder collaboration among different teams, as each team prefers its own "source of truth" [13] for data. This can lead to confusion and mistrust among teams, as they may have different versions of the same data. For example, if one team is using a different version of a customer database than another team, it can lead to discrepancies in customer information and hinder collaboration between the two teams. It also makes it difficult to share data across different systems and applications, which can further exacerbate the problem [14]. For example, an application deployed in a VMware host. Infrastructure team manages the VMware and application team manages the software. Each deployed their own monitoring tool and same data is being collected. This also results in deploying multiple different agents that consume the same OS resources for the same metrics – taking up storage, network bandwidth and compute.
- Tool/Vendor Lock-In: Normalized data stores are data repositories that follow a standard data model and structure to ensure data consistency and integrity. However, normalized data stores can also create data silos, which are isolated or disconnected data sources that prevent data integration and utilization across different platforms, systems, or applications. Data silos can lock customers into specific tools that are compatible or optimized for the normalized data stores. This can limit their choices and flexibility to use other tools that may offer better features, performance, or cost. Moreover, data silos can put their business at risk if the tools they rely on change or lose support in the future. For example, if a tool becomes obsolete, discontinued, or incompatible with the normalized data store, the customers may face data loss, migration, or transformation issues. Therefore, normalized data stores can create data silos that can have negative impacts on customers' options and business continuity.
- Missing actionable insight: To meet the changing demands of their businesses, enterprises have developed applications at various stages over time. These applications are monolithic in nature, meaning that they are composed of tightly coupled components that function as a single unit. To deliver a response to the end user, these applications must communicate with each other through complex interactions. However, each application also generates its own data, such as logs, traces and metrics, that capture its performance and behavior. These data are stored in separate application areas, making it difficult to correlate them and identify the root causes of issues. As a result, AIOps tool, which is supposed to provide a holistic view of the entire request-to-response process, cannot do so effectively. AIOps tool needs to have access to the data from all the applications and be able to link them together to create a comprehensive picture of the system. It is obvious that data silos restrict visibility across different verticals [4].

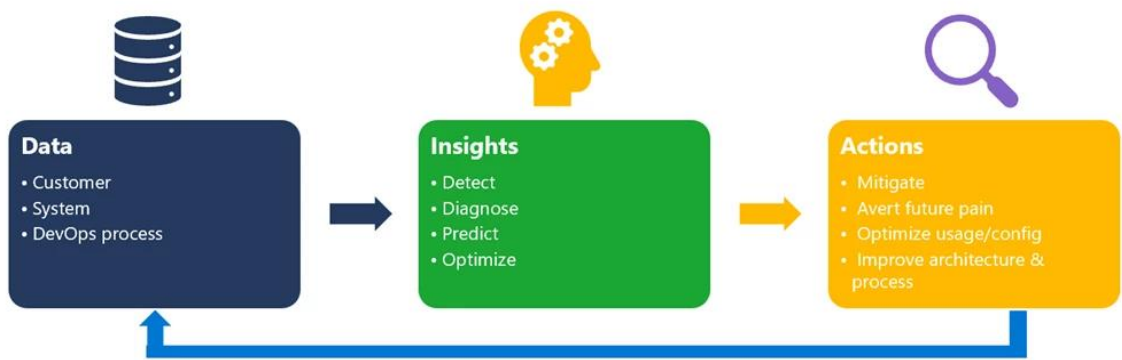

Fig 3. AIOps methodologies: Data, insights, and actions [5]

- Lessen the accuracy and integrity of your data: Data silos can lead to data duplication, inconsistency, and incompleteness. When data are stored in different places, they may not be updated or synchronized regularly, resulting in conflicting or missing information. This can affect the quality and reliability of data analysis and reporting. Data quality can be measured by various dimensions, such as accuracy, completeness, consistency, timeliness, relevance, and trustworthiness [6]. Data-backed decisions are only possible if the information in question is reliable and high-quality. Unfortunately, when data is isolated in a silo, it can lose relevance and eventually become unusable [7]. When data silos are formed, data fragmentation is inevitable, which can compromise data integrity. Recent research by Gartner shows that organizations believe poor data quality to be responsible for an average of $15 million per year in losses.
- Increased operational cost: Increase data leads increased infrastructure cost – storage cost, load on the network, replicating data over the network. AIOps highly relied on specialized hardware that can execute AI and ML models such as neural networks, decision trees, or clustering, with high speed and accuracy. Hardware systems that can host and run AI, ML, and MR applications or services, such as anomaly detection, root cause analysis, or automated actions, with high scalability and reliability. Specialized servers can be customized to have more processing cores, memory, storage, or network interfaces, depending on the workload requirements. Specialized servers can also be equipped with AI and ML engines or other accelerators, such as tensor processing units (TPUs) or neuromorphic chips, to boost the performance and efficiency of AI, ML, and MR techniques.

- Risks: Data silos can have negative impacts on the potential and value of data, as well as the exposure and vulnerability of data. Data silos can limit the potential and value of data by reducing the scope, quality, and accessibility of data for analysis, decision making, and innovation. Data silos can also increase the exposure and vulnerability of data by creating data security, privacy, or regulatory challenges. Data silos can make it difficult to protect, monitor, and audit data, as they may not comply with the data security, privacy, or regulatory requirements. Data silos can also increase the risk of data breaches, leaks, or losses, as they may not have adequate data backup, recovery, or encryption mechanisms. Therefore, data silos can hamper the benefits and opportunities of data and pose threats and challenges to data.

## 5 OVERCOME DATA SILOS

Here are some ways to overcome data silos.

- Centralized Data System: Data integration is a crucial aspect of developing microservices. Data integration is the process of combining data from different sources and systems into a unified and consistent view. In a centralized data system, several applications can read from and update a single data source. This can simplify the data integration process, as the data is stored and managed in a consistent and standardized way. However, a centralized data system can also have some drawbacks, such as scalability, performance, and reliability issues, as the data source can become a bottleneck or a single point of failure for the applications. Therefore, data integration in a centralized data system requires careful planning, design, and implementation, as well as monitoring, backup, and recovery mechanisms.

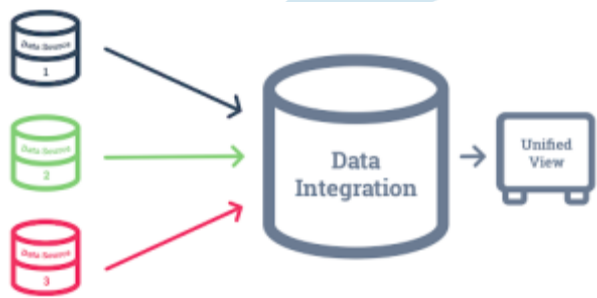

Fig 4. Integrated Systems [8]

- Standardize data formats to standardize tools: Standardizing the data formats used across the organization can have many benefits for data integration, quality, and usability. Data formats are the ways of representing and storing data in a file or a system. Different data formats may have different structures, syntaxes, and semantics that can affect how data can be accessed, processed, and exchanged. By using standardized and widely accepted open data formats, such as xml, json, csv, or sql, the organization can ensure that data can be easily and consistently communicated between different applications, platforms, or systems. This can reduce the complexity, cost, and error of data transformation and mapping processes. It can also improve the data quality and reliability by avoiding data loss, duplication, or inconsistency. Moreover, it can enhance the data usability and value by enabling data analysis, visualization, and sharing across the organization. One of the outcomes of this discussion is the need to standardize the tools and use more open-source frameworks that can leverage vendor-independent or vendor-neutral APIs. This can help the organization avoid being locked in by a specific vendor or tool that may change or lose support in the future. For example, if the organization follows the opentelemetry framework [18] while developing codes, it can benefit from a wide range of AIOps tools that can monitor and communicate with those codes. The opentelemetry framework is an open-source and vendor-neutral API that provides a standard way of collecting and exporting telemetry data, such as metrics, traces, and logs, from different sources and systems. By using this framework, the organization can ensure the compatibility and interoperability of its codes and tools across different platforms and environments.
- Develop data culture: Data analytics is becoming a pervasive and essential aspect of modern organizational life. Organizations are investing trillions to become more data-driven, but only 8% successfully scale analytics to get value from their data [20]. This means that having a data culture that values and uses data effectively is more and more important. In 2018, a quarterly report of McKinsey has been published and they interviewed several C-level executives. They found that Data culture is not something that you can bring in from outside or force on your organization. It is also not something that you can isolate or separate from your core activities. You build a data culture by going beyond relying on experts and isolated projects, and aiming for deep engagement with your business, strong demand from your employees, and clear alignment with your purpose, so that data can enable your operations rather than constrain them [17]. The use of collaborative technology is often blocked by a behavior barrier - the problem is not the technology, but the culture that needs to change [16]. The organization should foster a strong data culture and governance. The data office should establish a guideline on how to ingest, extract, transform, duplicate, and purge data. The management should ensure the compliance and consistency of data across the organization. Organizations need a strong data retention strategy and audit to avoid accumulating and storing data, logs, metrics that are useless for AIOps. Therefore, organizations should have a clear and consistent policy and process to determine how long and why they should keep or delete data, logs, metrics that are related to AIOps.
- Integrated Systems and Data Integration: The organization aims to build integrated systems that can communicate and cooperate with each other. To achieve this, the organization should have a plan to migrate its segregated monolith applications to microservices. Monolith applications are applications that are built as a single and indivisible unit, where all the components and functionalities are tightly coupled and dependent on each other. Microservices are applications that are built as a collection of small and independent units, where each unit has a specific function and can communicate with other units through well-defined interfaces. By migrating to microservices, the organization can benefit from increased scalability, flexibility, reliability, and performance of its systems. It can also reduce the complexity, cost, and risk of developing, testing, and deploying its systems. However, migrating to microservices also requires careful planning, design, and implementation, as it involves breaking down the monolith applications into smaller units, defining the interfaces and protocols for communication, and managing the dependencies and coordination among the units.

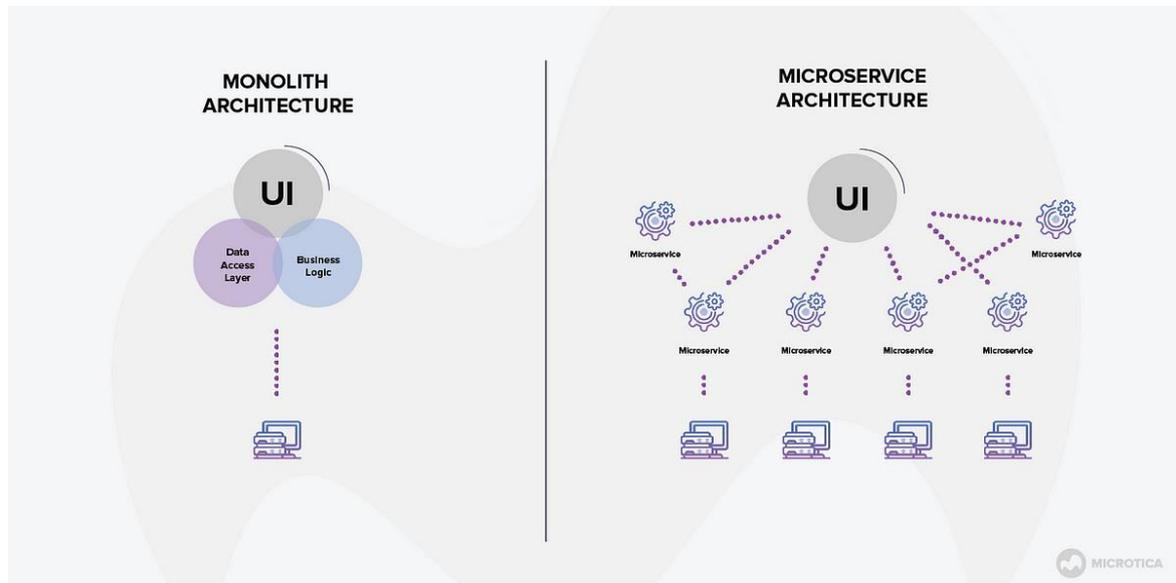

Fig 5: Monolith to Microservice [19]

## 6 SUMMARY

Data silos are isolated repositories of data that are not easily accessible or shared by other systems or departments within an organization. Data silos can hinder the efficiency, accuracy, and collaboration of data-driven decision making. However it is inevitable. Therefore, the aim of the organization is to minimize the silos, collaboration and communication, develop data culture and data governance, breaking down monolith and legacy application. By doing so, the organization can improve the quality, consistency, and availability of data across different platforms and teams. This can lead to better insights, innovation, and customer satisfaction which are key for any AIOps to be successful.

## 7 ACKNOWLEDGMENTS

I would like to thank anonymous reviewers for the comments and suggestions.

## 8 REFERENCES

[1] Cisco. "Global AI Readiness Index." 2020, https://www.cisco.com/c/dam/m/en_us/solutions/ai/readiness-index/documents/cisco-global-ai-readiness-index.pdf.

[2] Patel, Jayesh. "Bridging Data Silos Using Big Data Integration." International Journal of Database Management Systems, vol. 11, no. 3, 2019, pp. 01-06.

[3] Baker, Chris. "What Are Data Silos and How Can You Eliminate Them?" Adobe, 9 Apr. 2020, https://business.adobe.com/blog/basics/what-are-data-silos-and-how-can-you-eliminate-them.

[4] Patel, Jayesh. "Overcoming Data Silos Through Big Data Integration." International Journal of Computer Science and Technology, vol. 3, no. 1, 2019.

[5] Russinovich, Mark. "Advancing Azure Service Quality with Artificial Intelligence: Aiops." Microsoft Azure Blog, 11 May 2023, https://azure.microsoft.com/en-us/blog/advancing-azure-service-quality-with-artificial-intelligence-aiops/.

[6] Carruthers, Andrew. "Breaking Data Silos." Data Science for Business and Decision Making, by Luiz Paulo Fávero et al., Apress, 2019, pp. 19-36.

[7] Lin, Pohan. "What Are Data Silos? How to Interconnect Data and Increase Efficiency." G2, https://learn.g2.com/data-silos.

[8] Kirwan, Kelly. "How to Fix Data Silos & Unlock Your Data's Full Value." Segment, 1 Dec. 2021,

https://segment.com/blog/fix-data-silos/.

[9] Lin, Pohan. "Solutions to Common Data Silo Problems." IEEE Computer Society, 1 Sep. 2023, [1](https://www.computer.org/publications/tech-news/trends/solutions-to-common-data-silo-problems/).

[10] Quixy Editorial Team. "Data Silos: Breaking Down the Basics!" Quixy, 10 Aug. 2023, [1](https://quixy.com/blog/data-silos/).

[11] Sabharwal, Navin, and Gaurav Bhardwaj. "What Is AIOps?" Hands-on AIOps, Apress, 2022, pp. 1-14. doi: 10.1007/978-1-4842-8267-0_1.

[12] Moore, Susan. "How to Create a Business Case for Data Quality Improvement." Gartner, 19 June 2018.

[13] Oppermann, Artem. "What Are Data Silos?" Built In, 6 Nov. 2023.

[14] Skinner, Matt. "What Are Data Silos and How Can You Eliminate Them?" Adobe, 17 Oct. 2022.

[15] Chawla, Vishal. "Why You Need To Break Data Silos To Build Powerful AI Systems." Analytics India Magazine, 11 June 2020.

[16] Cromity, Jamal, and Ulla De Stricker. "Silo persistence: It's not the technology, it's the culture!." New Review of Information Networking 16.2 (2011): 167-184.

[17] Díaz, Alejandro, Kayvaun Rowshankish, and Tamim Saleh. "Why data culture matters." McKinsey Quarterly 3.1 (2018): 36-53.

[18] Thakur, Aadi, and M. B. Chandak. "A review on opentelemetry and HTTP implementation."

[19] Miteva, Sara. "Why Transition from Monolith to Microservices?" Microtica, 3 Mar. 2020.

[20] Bisson, Peter, et al. "Breaking Away: The Secrets to Scaling Analytics." Apr. 2018.

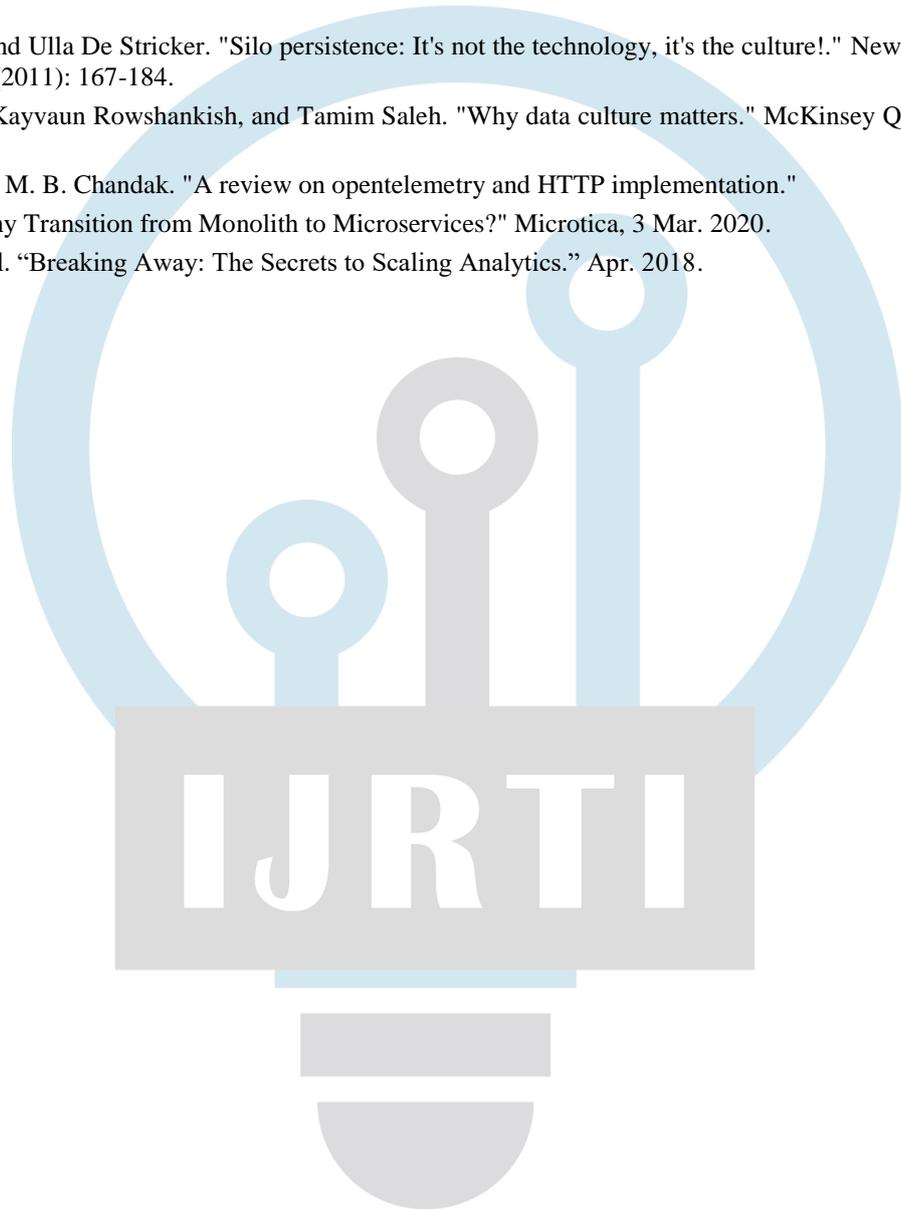